\documentclass[conference]{IEEEtran}
\IEEEoverridecommandlockouts

\usepackage{cite}
\usepackage{amsmath,amssymb,amsfonts}
\usepackage{algorithmic}
\usepackage{graphicx}
\usepackage{colortbl, hhline}
\usepackage{textcomp}
\usepackage{xcolor}
\usepackage{mathtools}
\usepackage{braket}

\usepackage{tikz}
\usetikzlibrary{arrows}

\def\BibTeX{{\rm B\kern-.05em{\sc i\kern-.025em b}\kern-.08em
    T\kern-.1667em\lower.7ex\hbox{E}\kern-.125emX}}

\newcommand{\verteq}{\rotatebox{90}{$\,=$}}

\begin{document}

\title{A Game-Theoretic Quantum Algorithm \\ for Solving Magic Squares \\

\thanks{This manuscript has been authored by UT-Battelle, LLC under Contract No. DE-AC05-00OR22725 with the U.S. Department of Energy. The United States Government retains and the publisher, by accepting the article for publication, acknowledges that the United States Government retains a non-exclusive, paid-up, irrevocable, world-wide license to publish or reproduce the published form of this manuscript, or allow others to do so, for United States Government purposes. The Department of Energy will provide public access to these results of federally sponsored research in accordance with the DOE Public Access Plan (http://energy.gov/downloads/doe-public-access-plan).}
}

\author{\IEEEauthorblockN{Sarah Chehade}
\IEEEauthorblockA{\textit{Quantum Information Sciences Section} \\
\textit{Oak Ridge National Laboratory}\\
Oak Ridge, USA \\
sarahschehade4@gmail.com}
\and
\IEEEauthorblockN{Andrea Delgado}
\IEEEauthorblockA{\textit{Physics Division} \\
\textit{Oak Ridge National Laboratory}\\
Oak Ridge, USA \\
delgadoa@ornl.gov}
\and
\IEEEauthorblockN{Elaine Wong}
\IEEEauthorblockA{\textit{Computer Science and Math Division} \\
\textit{Oak Ridge National Laboratory}\\
Oak Ridge, USA \\
wongey@ornl.gov}
}

\maketitle

\begin{abstract}
Variational quantum algorithms (VQAs) offer a promising near-term approach to finding optimal quantum strategies for playing non-local games. These games test quantum correlations beyond classical limits and enable entanglement verification. In this work, we present a variational framework for the Magic Square Game (MSG), a two-player non-local game with perfect quantum advantage. We construct a value Hamiltonian that encodes the game's parity and consistency constraints, then optimize parameterized quantum circuits to minimize this cost. Our approach builds on the stabilizer formalism, leverages commutation structure for circuit design, and is hardware-efficient. Compared to existing work, our contribution emphasizes algebraic structure and interpretability. We validate our method through numerical experiments and outline generalizations to larger games. 
\end{abstract}


\section{Introduction}
\label{sec:intro}

Non-local games (NLGs) are interactive tasks in which non-communicating players attempt to win against a referee using a shared quantum state. They are central to quantum information science (QIS), with applications in quantum cryptography~\cite{coladangelo2021,culf2022monogamy,Zhen2023}, device-independent protocols~\cite{Cleve2004, ito2008, Kalai2023}, and foundational tests of quantum mechanics. NLGs extend the concept of Bell inequalities, operationalizing the notion of quantum advantage as the ability to win games with probabilities exceeding classical limits.

The violation of Bell inequalities reveals the nonclassical nature of quantum correlations. In a Bell scenario, the set of classically achievable correlations obeys a convex polytope defined by linear inequalities. Quantum mechanics allows for correlations that lie outside this polytope, manifesting as violations of the corresponding Bell inequalities. These violations not only serve as entanglement witnesses but also underlie the security of quantum key distribution and the trustworthiness of quantum devices in a semi-device-independent framework.

In this work, we explore a variational quantum algorithm (VQA)-style approach to discover optimal strategies for the Magic Square Game (MSG)~\cite{Mermin1990b, Peres1990}, a 2-player non-local binary constraint game that exhibits perfect quantum advantage. While optimal strategies for small games can often be derived analytically, variational methods offer a scalable alternative that is well suited for near-term quantum hardware. The idea is to encode the game’s winning conditions into a value Hamiltonian and use a parameterized quantum circuit to optimize the measurement strategy. This enables hardware-friendly, data-driven discovery of quantum strategies in non-local settings.

The MSG is particularly compelling: it highlights the distinction between classical and quantum logic through parity constraints, allowing quantum players to win with 100$\%$ success using only three shared Bell pairs and appropriately chosen measurements. A key structural insight of our work is the role of commutativity in enforcing these game constraints. In any finite-dimensional quantum model of a non-local game, a fundamental requirement is that observables associated with each player commute. This ensures that measurements correspond to local operations on separate Hilbert spaces, preserving the no-signaling condition. We found that this commutativity is not merely a technical requirement for defining the game, but plays a constructive role in strategy optimization—enabling independent, variational updates to Alice's and Bob’s observables while preserving game consistency.

Variational approaches to learning quantum strategies for non-local games have recently gained attention. Notably, Ref.~\cite{furches2023variational} formulates the problem of learning quantum strategies using variational circuits, exploring training techniques to approximate optimal strategies across multiple games. While their work provides an excellent overview of parameter landscapes and optimization schemes, our contribution differs in both scope and emphasis. We focus on a single game (MSG), dissecting its stabilizer structure in detail, and uncovering how the algebraic properties of commuting measurement operators are exploited to ensure constraint satisfaction.

Similar strategies also appear in previous works. In Ref.~\cite{bharti2019teachaiplaybell}, reinforcement learning is applied to learn optimal measurement settings and states for Bell scenarios. While they use reinforcement learning as the optimization driver, our work relies on deterministic gradient-based methods within the VQA framework. Moreover, their focus is on more generic inequalities such as CHSH and bilocality, while our approach addresses games with algebraically constrained outputs.

In Ref.~\cite{Sheffer2022}, the authors implement non-local games using shallow circuits on real quantum hardware. Their work emphasizes experimental realization and hardware compatibility. Our approach is complementary: rather than adapting a fixed measurement strategy to noisy hardware, we variationally discover the strategy that satisfies the game rules, with the potential to generalize to larger games or self-testing scenarios.

In this work, we present a variational framework for discovering optimal quantum strategies for the MSG, leveraging the stabilizer formalism and modern VQA techniques. Our approach begins with a principled construction of measurement operators that encode the game's parity constraints as stabilizer-like observables. We show that the commutation structure of these operators—within each player's subsystem—ensures compatibility with the no-signaling condition while also enabling effective variational optimization. By minimizing a value Hamiltonian that encodes the game’s success conditions, we are able to recover the perfect quantum strategy, achieving the maximum game value. Through exhaustive numerical validation, we demonstrate that our method correctly enforces both parity and intersection consistency across all possible game inputs. In contrast to previous works that explore variational strategies for a range of NLGs or emphasize experimental implementation, our contribution lies in dissecting the algebraic underpinnings of the MSG and demonstrating how a variationally trained quantum circuit can learn and enforce the game’s exact constraints from first principles.

\section{Theoretical Framework}
\label{sec:theory}

\subsection{Non-local Games and Quantum Strategies}

\begin{figure}
\centering
\includegraphics[scale=1]{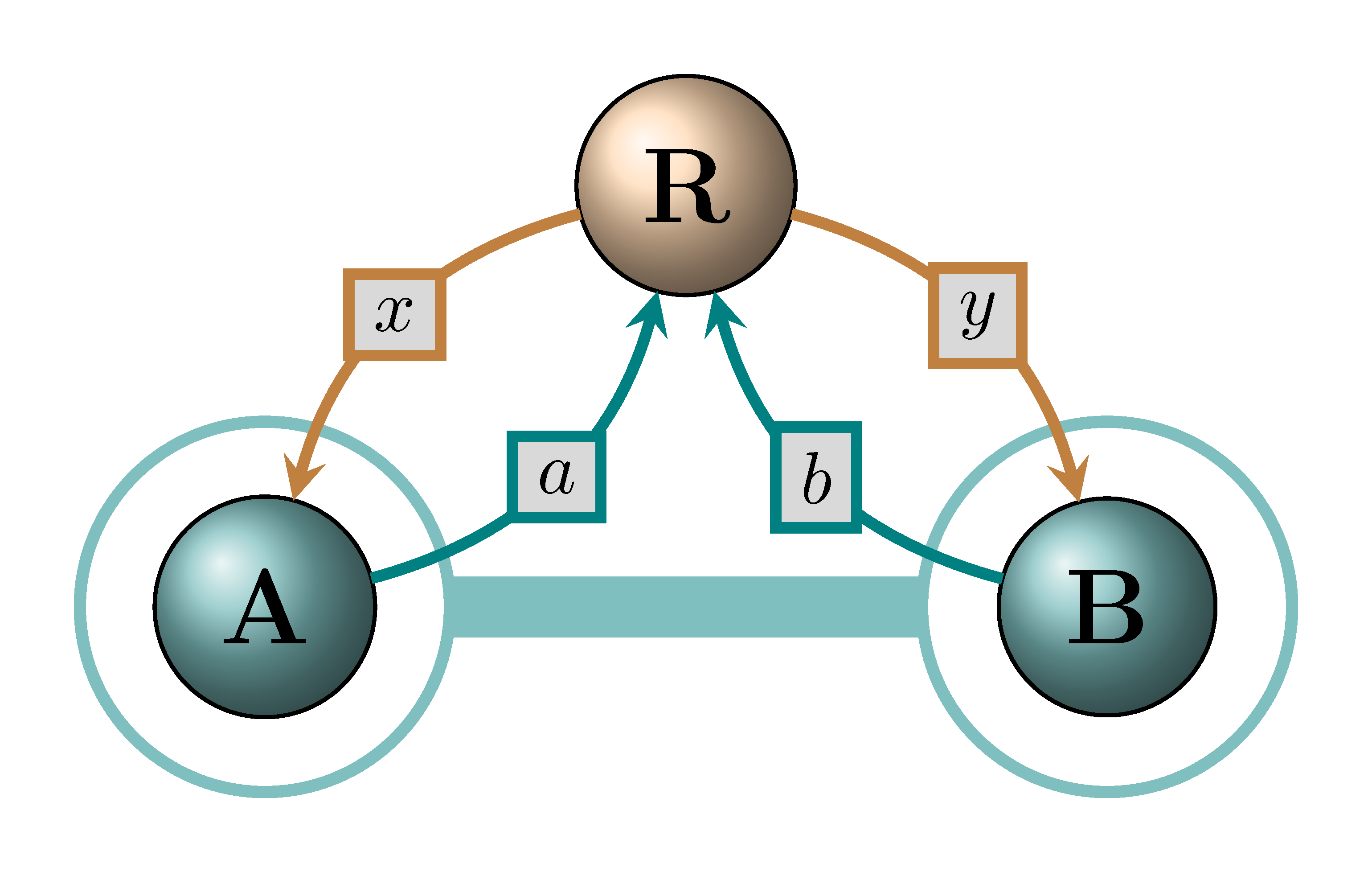}
\caption{A game with a referee R and two players, (A)lice and (B)bob, sharing an entangled state, indicated symbolically by the light teal band. Upon receiving questions $x$ and $y$ from the referee, they return answers $a$ and $b$, respectively. Prior to the game starting, all parties know the rules of the game and the strategy they will use, but will not be able to communicate during the game.}
\label{fig:players}
\end{figure}

A non-local game (NLG) is defined as an interactive task between a referee and multiple players who share an entangled state and can discuss game strategy, but cannot communicate after the game starts. Upon receiving inputs $x, y$ from a predefined set, each player must return outputs $a, b$ from a different predefined set (see Figure~\ref{fig:players}). The players win or lose the game based on a rule $r(x,y,a,b)$ specified by the referee, who selects inputs $(x,y)$ from a distribution $\pi(x,y)$. In the quantum model, each player holds a part of a shared quantum state $\ket{\psi} \in \mathcal{H}_A \otimes \mathcal{H}_B$ and responds to input $x$ (or $y$) by measuring an observable from a set of Hermitian operators $\{A_x^a\}, \{B_y^b\}$ acting on their subsystem.

The quantum value of the game is defined as
\begin{equation}
    \omega_q := \sum_{x,y} \pi(x,y) \sum_{a,b} r(x,y,a,b) \bra{\psi} A_x^a \otimes B_y^b \ket{\psi},
\end{equation}
which computes the probability at which the players win the game for the given strategy.

\subsection{The Magic Square Game}
We focus on the $3\times 3$ Magic Square Game (MSG), a two-player NLG where players are tasked with filling a $3\times3$ grid with entries in $\{\pm1\}$ that satisfy specific row and column parity constraints. In one version of the game, Alice is given a row index $r \in \{0,1,2\}$ and must produce three bits such that the product of entries in that row is $+1$, while Bob is given a column index $c \in \{0,1,2\}$ and must produce three bits such that the product in that column is $-1$. Figure~\ref{fig:msg} shows how such a construction can be understood as a square with entries in the special unitary group SU(4), and their corresponding row and column products. They win the round if their answers agree on the overlapping cell $(r,c)$. No classical deterministic or probabilistic strategy can win all nine combinations of inputs, and the classical value is bounded by $\omega_c = 8/9$. Quantumly, a perfect strategy exists with success probability $\omega_q = 1$, achievable by using shared entanglement and measurements that satisfy certain algebraic consistency conditions.

\begin{figure}[ht]
\centering
\renewcommand{\arraystretch}{2}
\begin{tabular}{cccc}
    \hhline{---~}
    \cellcolor{red!20}  $I\otimes Z$ & \cellcolor{red!20}  $Z\otimes I$                & \cellcolor{red!20}  $Z\otimes Z$  & $=I\otimes I$\\ \hhline{---~}
    \cellcolor{red!20}  $X\otimes I$ & \cellcolor{red!20}  $I\otimes X$ & \cellcolor{red!20}  $X\otimes X$ & $=I\otimes I$ \\ \hhline{---~}
    \cellcolor{red!20}  $X\otimes Z$ & \cellcolor{red!20}  $Z\otimes X$                & \cellcolor{red!20}  $Y\otimes Y$  & $=I\otimes I$ \\ \hhline{---~}
    \verteq & \verteq & \verteq & \\[-5mm]
    $I\otimes I$ & $I\otimes I$ & $-I\otimes I$ &    
\end{tabular}
\caption{This shows a prototypical example of a `strategy' for the magic square game consisting of observables on two qubits that encode some game constraints (i.e., product of entries equals the values printed outside of the square in their respective row/column), on which both players agree before the game begins. The observables consists of elements from the Pauli group $\{X, Y, Z, I\}$. Upon receiving their question row/column, the elements in the corresponding row (for player A) or column (for player B) are used to construct their measurement operators, which enables the computation of the answers that they will give. In section \ref{sec:construction}, we will illustrate a general set up for three-qubit systems, which more easily enables Hamiltonian construction for the variational algorithm.}
\label{fig:msg}
\end{figure}

\subsection{Operator Construction and Strategy Encoding}
\label{sec:construction}

\begin{figure}
\centering
\includegraphics[scale=1.2]{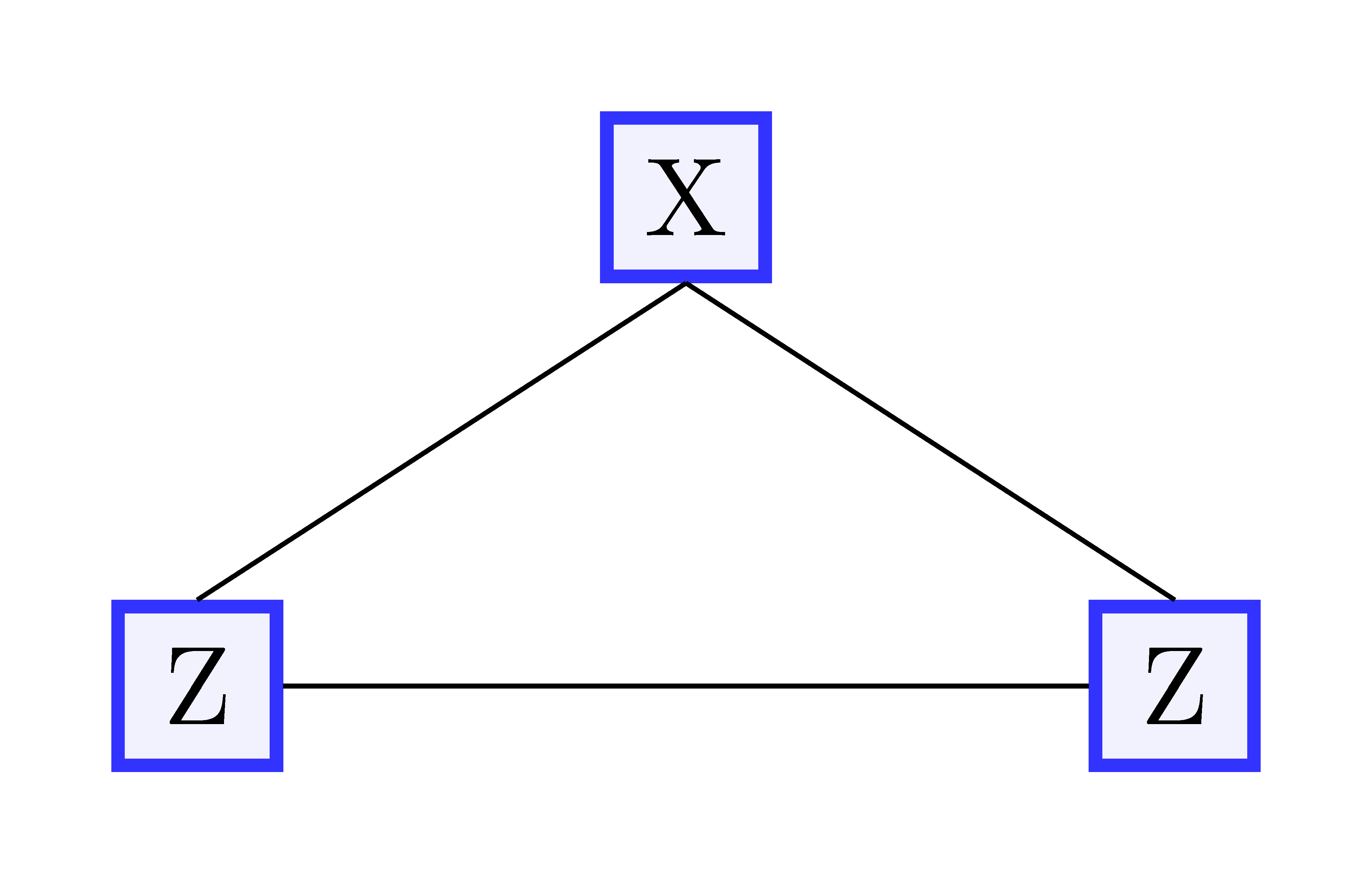}
\includegraphics[scale=1.2]{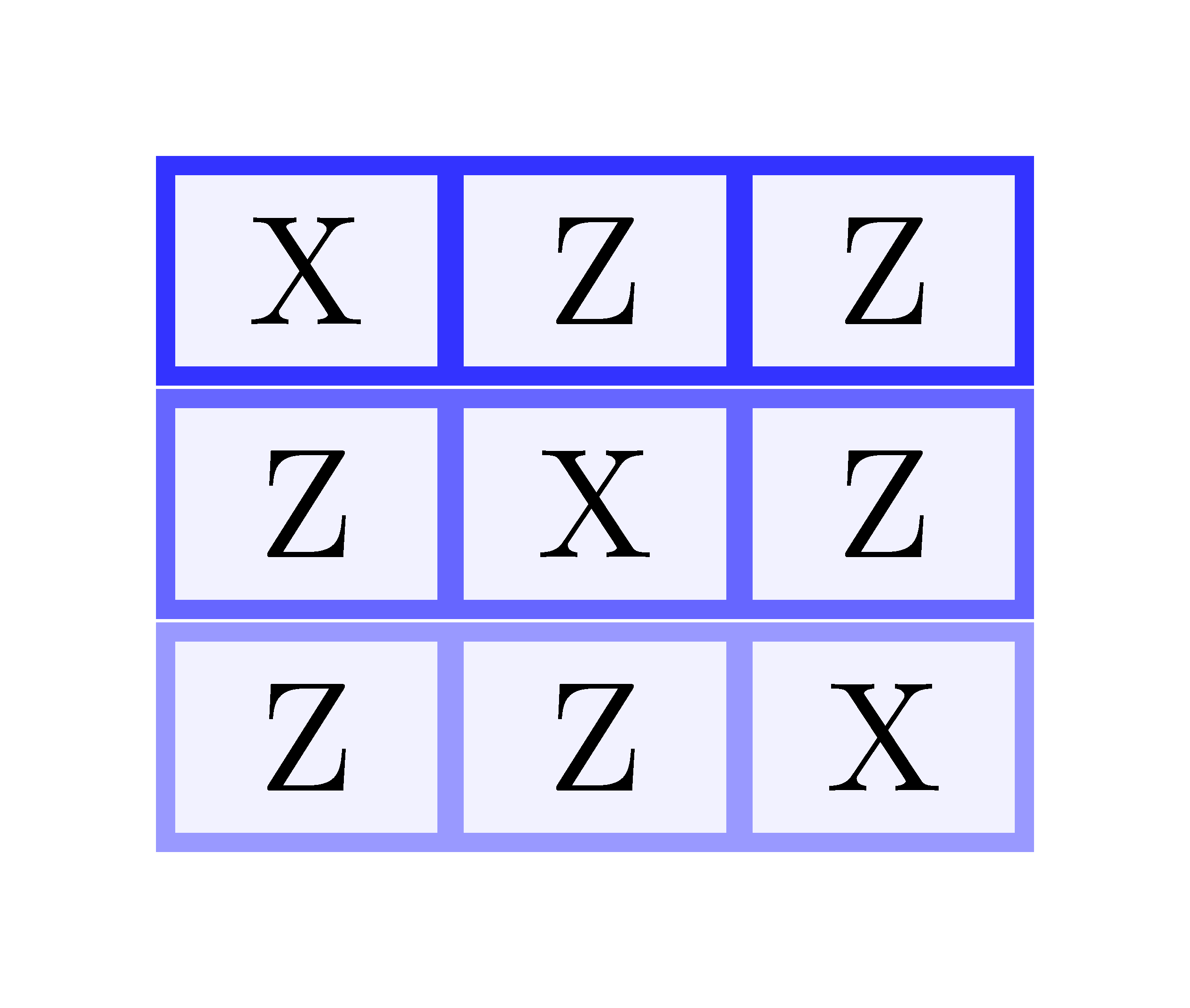}
\caption{Mermin's GHZ game~\cite{Mermin1990b} is typically played between a referee and \textit{three} players and samples questions from a pre-defined set. Referee accepts answers under certain parity constraints. The stabilizers for the triangle graph state are the candidates for observables that satisfy the required constraints, which can be achieved by taking the product of the stabilizer operators. Our choice of adapting the two-player game to a three-qubit system is justified from the construction of a Hamiltonian that can encode parity constraints.}
\label{fig:mermin}
\end{figure}

The key to implementing a quantum strategy lies in defining a set of fixed observables $\{A_i\}, \{B_j\}$—one per row for Alice and one per column for Bob—that are used to construct \emph{projectors} encoding the game constraints. Each observable acts on a three-qubit subsystem and is constructed from tensor products of Pauli matrices:
\begin{equation}
    A_i = P_i^{(1)} \otimes P_i^{(2)} \otimes P_i^{(3)}, \qquad B_j = Q_j^{(1)} \otimes Q_j^{(2)} \otimes Q_j^{(3)},
    \label{eq:abobs}
\end{equation}
where $P_i^{(k)}, Q_j^{(k)} \in \{X, Z\}$. For all operators, we impose the structural rule that each contains exactly one Pauli-X and two Pauli-Z operators, ensuring compatibility with the game’s parity constraints (see Figure~\ref{fig:mermin}). These operators are Hermitian and unitary, with eigenvalues in $\{\pm1\}$, and are used to define projectors:
\begin{equation}
    \Pi_{i,j}^{\text{win}} = \frac{1}{2} \left( \mathbb{I} + A_i \otimes B_j \right)
\end{equation}
which project onto winning subspaces for input pair $(i,j)$.

These operators are not the direct measurement operators; rather, they are fixed, observable-valued labels defining the game structure. The actual measurements performed by the players are constructed from these operators via parameterized unitaries, as we explain next.

\subsection{Commutation Structure and Consistency}

A central feature of our approach is that the fixed operators \( A_i \) and \( B_j \) are mutually commuting within each player’s set: \( [A_i, A_{i'}] = 0 \) and \( [B_j, B_{j'}] = 0 \) for all $i\neq i'$ and $j\neq j'$, ensuring local compatibility. However, operators across players generally do not commute and may anticommute, especially when corresponding to overlapping cells. The consistency rule of the game demands that Alice’s and Bob’s outputs agree on these intersections, and we enforce this condition implicitly through optimization. Specifically, we verify that for each overlapping cell $(i,j)$, the equality of measurement outcomes is preserved:
\begin{equation}
    \bra{\psi} \tilde{A}_{i} \otimes \tilde{B}_{j} \ket{\psi} \to +1.
\end{equation}

Empirically, we observe that the expectation values of all terms in the Hamiltonian converge to $+1$ under training, resulting in a total value of $-9$, which is the minimal eigenvalue of $H$ and corresponds to perfect game success. The evolution of commutators and operator overlaps during training supports the interpretation that our variational circuit learns a quantum strategy consistent with both the physical locality and the algebraic structure of the MSG.

\begin{table}[ht]
    \caption{List of projectors $\Pi_{i,j}^{\text{win}}$ used in constructing the value Hamiltonian for the Magic Square Game. Each projector is formed from the tensor product of a row operator $A_i$ and column operator $B_j$, where $A_i, B_j \in \{X, Z\}^{\otimes 3}$. The complete value Hamiltonian is given by $H = -\sum_{i,j} A_i \otimes B_j$.}
    \label{tab:projectors}
    \centering
    \renewcommand{\arraystretch}{1.4}
    \begin{tabular}{ccc}
        \hline
        \textbf{Row $i$} & \textbf{Column $j$} & \textbf{Projector} \\
        \textbf{(Alice)} & \textbf{(Bob)} & $\Pi_{i,j}^{\text{win}} = \frac{1}{2}(\mathbb{I} + A_i \otimes B_j)$ \\
        \hline
        $A_0 = Z \otimes Z \otimes X$ & $B_0 = X \otimes Z \otimes Z$ & $\frac{1}{2}(\mathbb{I} + A_0 \otimes B_0)$ \\
        $A_0 = Z \otimes Z \otimes X$ & $B_1 = Z \otimes X \otimes Z$ & $\frac{1}{2}(\mathbb{I} + A_0 \otimes B_1)$ \\
        $A_0 = Z \otimes Z \otimes X$ & $B_2 = Z \otimes Z \otimes X$ & $\frac{1}{2}(\mathbb{I} + A_0 \otimes B_2)$ \\
        $A_1 = X \otimes Z \otimes Z$ & $B_0 = X \otimes Z \otimes Z$ & $\frac{1}{2}(\mathbb{I} + A_1 \otimes B_0)$ \\
        $A_1 = X \otimes Z \otimes Z$ & $B_1 = Z \otimes X \otimes Z$ & $\frac{1}{2}(\mathbb{I} + A_1 \otimes B_1)$ \\
        $A_1 = X \otimes Z \otimes Z$ & $B_2 = Z \otimes Z \otimes X$ & $\frac{1}{2}(\mathbb{I} + A_1 \otimes B_2)$ \\
        $A_2 = Z \otimes X \otimes Z$ & $B_0 = X \otimes Z \otimes Z$ & $\frac{1}{2}(\mathbb{I} + A_2 \otimes B_0)$ \\
        $A_2 = Z \otimes X \otimes Z$ & $B_1 = Z \otimes X \otimes Z$ & $\frac{1}{2}(\mathbb{I} + A_2 \otimes B_1)$ \\
        $A_2 = Z \otimes X \otimes Z$ & $B_2 = Z \otimes Z \otimes X$ & $\frac{1}{2}(\mathbb{I} + A_2 \otimes B_2)$ \\
        \hline
    \end{tabular}
\end{table}

\section{Methods}

We implement a variational quantum algorithm to find an optimal quantum strategy for the $3\times 3$ MSG. Our method follows a hybrid quantum-classical optimization scheme in which parameterized circuits define the players' measurements, and a classical optimizer adjusts parameters to minimize the game cost function. The approach allows us to implicitly enforce the consistency and compatibility conditions required for a quantum advantage in NLGs.

We begin by defining a fixed set of initial observables $\{A_{i}\}$ for Alice and $\{B_{j}\}$ for Bob, each acting on a 3-qubit subsystem, and each operator includes exactly two Z and one X. This structure mirrors the operator assignments in the original Mermin-Peres magic square and enforces the required anti-commutation properties necessary for the game's nonclassical constraints

We verify analytically and numerically that these operators satisfy the required row and column parity conditions and exhibit the desired anti-commutation and commutation behavior on the intersections~\cite{arkhipov2012extending}.

\subsection{State Preparation}
The quantum strategy begins with a six-qubit entangled state composed of three Bell pairs, one for each corresponding pair of qubits between Alice and Bob:

\begin{equation}
    \ket{\psi} = \ket{\Phi^+}_{A_0 B_0} \otimes \ket{\Phi^+}_{A_1 B_1} \otimes \ket{\Phi^+}_{A_2 B_2},
\end{equation}

where
\begin{equation}
    \ket{\Phi^+} = \frac{1}{\sqrt{2}}(\ket{00} + \ket{11}).
\end{equation}

Each Bell pair ensures maximal correlation between the qubits held by Alice and Bob at the same index. The full state thus provides the entanglement structure required for non-local correlations that violate classical bounds.

To enforce the parity constraints of the MSG, each player must be able to assign values to three entries (i.e., bits) in a row or column. Therefore, each player must hold at least three qubits. Specifically, Alice's measurement of a row corresponds to a three-bit string subject to an \emph{odd parity} constraint, while Bob’s column measurement must satisfy an \emph{even parity} constraint. The output of the measurement must collapse the player’s qubit register into an eigenstate of the corresponding parity-check operator. Since these operators act nontrivially on three qubits, a three-qubit local Hilbert space per player is the minimal requirement for representing their full local strategy.

Compared to the dual-optimization method in Ref.~\cite{furches2023variational}, our single-phase approach avoids alternating optimization and directly minimizes the value Hamiltonian using fixed entanglement and parameterized observables. This leads to faster convergence and simpler implementation for small games.

\subsection{Operator Encoding and Value Hamiltonian}

Each row \( i \in \{0,1,2\} \) for Alice and column \( j \in \{0,1,2\} \) for Bob is associated with a parity-check operator \( A_i \in \{X,Z\}^{\otimes 3} \) and \( B_j \in \{X,Z\}^{\otimes 3} \), respectively. These fixed observables encode the parity constraint algebraically. The full game behavior is captured by the value Hamiltonian:

\begin{equation}
    H = -\sum_{i,j=0}^{2} A_i \otimes B_j,
\end{equation}

where each term evaluates to $+1$ if Alice and Bob’s answers for input pair $(i,j)$ are valid and consistent, and $-1$ otherwise. Thus, the ground state energy of $H$ corresponds to perfect game performance with score $-9$.

\subsection{Parameterized Measurement Operators}

To enable variational learning of a strategy, we introduce local unitaries \( U_i(\theta) \) for Alice and \( V_j(\phi) \) for Bob that rotate the fixed operators \( A_i \), \( B_j \) into trainable measurement bases:

\begin{equation}
    \tilde{A}_i = U_i^\dagger(\theta) A_i U_i(\theta), \quad \tilde{B}_j = V_j^\dagger(\phi) B_j V_j(\phi).
    \label{eq:abtilde}
\end{equation}

This construction allows for expressive control over the measurement basis via tunable parameters, while preserving the parity-check structure encoded in $A_i$ and $B_j$. Each unitary acts on a three-qubit register and is implemented using the \texttt{StronglyEntanglingLayers} template in \textit{PennyLane}~\cite{bergholm2022pennylane}, a hardware-efficient ansatz composed of:

\begin{itemize}
    \item a layer of single-qubit rotations in the $X$ and $Y$ directions,
    \item followed by a fixed pattern of entangling gates (CNOT or CZ), and
    \item repeated for a fixed number of layers (we use 3).
\end{itemize}

These unitaries are fully differentiable and are optimized during training. The structure of the ansatz ensures full expressibility over the three-qubit local Hilbert space, while preserving locality and gate depth constraints suitable for near-term devices.

\subsection{Cost Function}

Our cost function is designed to penalize deviations from the perfect strategy value:

\begin{equation}
    \mathcal{L}(\theta, \phi) = \bra{\psi} \left( \sum_{i,j=0}^{2} \tilde{A}_i \otimes \tilde{B}_j \right) \ket{\psi}.
    \label{eq:cost}
\end{equation}

Minimizing $\mathcal{L}$ is equivalent to maximizing the expected score of the Magic Square Game. The global minimum of $-9$ corresponds to a perfect strategy where all expectation values $\braket{\tilde{A}_i \otimes \tilde{B}_j} = +1$. Because the measurement operators $\tilde{A}_i$ and $\tilde{B}_j$ are constructed from unitarily rotated projectors, the optimization implicitly enforces the consistency and parity constraints of the game.

\subsection{Training Procedure}

We optimize the parameters $\theta$ and $\phi$ using the Adam optimizer with an initial learning rate of $0.1$. Parameters are initialized from a standard normal distribution and reshaped to fit the layered circuit architecture. Each optimization step computes the gradient of the cost function via automatic differentiation using \texttt{PennyLane}’s \texttt{autograd} backend. The VQE loop is executed for 200 iterations, during which we track

\begin{enumerate}
    \item the convergence of the loss function to $-9$,
    \item the norm of the parameter updates,
    \item the commutators $[\tilde{A}_i, \tilde{A}_{i'}]$ and $[\tilde{B}_j, \tilde{B}_{j'}]$, and
    \item the consistency condition at overlapping cells.
\end{enumerate}

Empirically, the optimization consistently converges to the theoretical minimum, validating both the ansatz and the initial operator construction.

\section{Results}
\label{sec:results}

To evaluate the quality of the learned quantum strategy for the MSG, we trained a variational quantum circuit to minimize the expectation value of a Hamiltonian encoding the game constraints. This Hamiltonian is constructed from nine tensor products of the form $A_{i} \otimes B_{j}$, where $A_{i}$ and $B_{j}$ are the three-qubit Pauli observables for Alice and Bob, respectively (Table~\ref{tab:projectors}). The cost function, defined as the expectation value $\braket{H}$, reaches its minimum when the strategy maximally satisfies the game's non-local correlations. In all optimization runs, we observe convergence to a game value close to the theoretical optimum of 9, indicating that the trained quantum circuit captures the structure of the wining quantum strategy.

To interpret the behavior of the strategy more precisely, in what follows, we examine several aspects of the results during (cost function and parameter stability) and after training (convergence and constraint verification).

\subsection{Cost Function Convergence}

\begin{figure}[ht]
    \centering
    \includegraphics[width=0.5\textwidth]{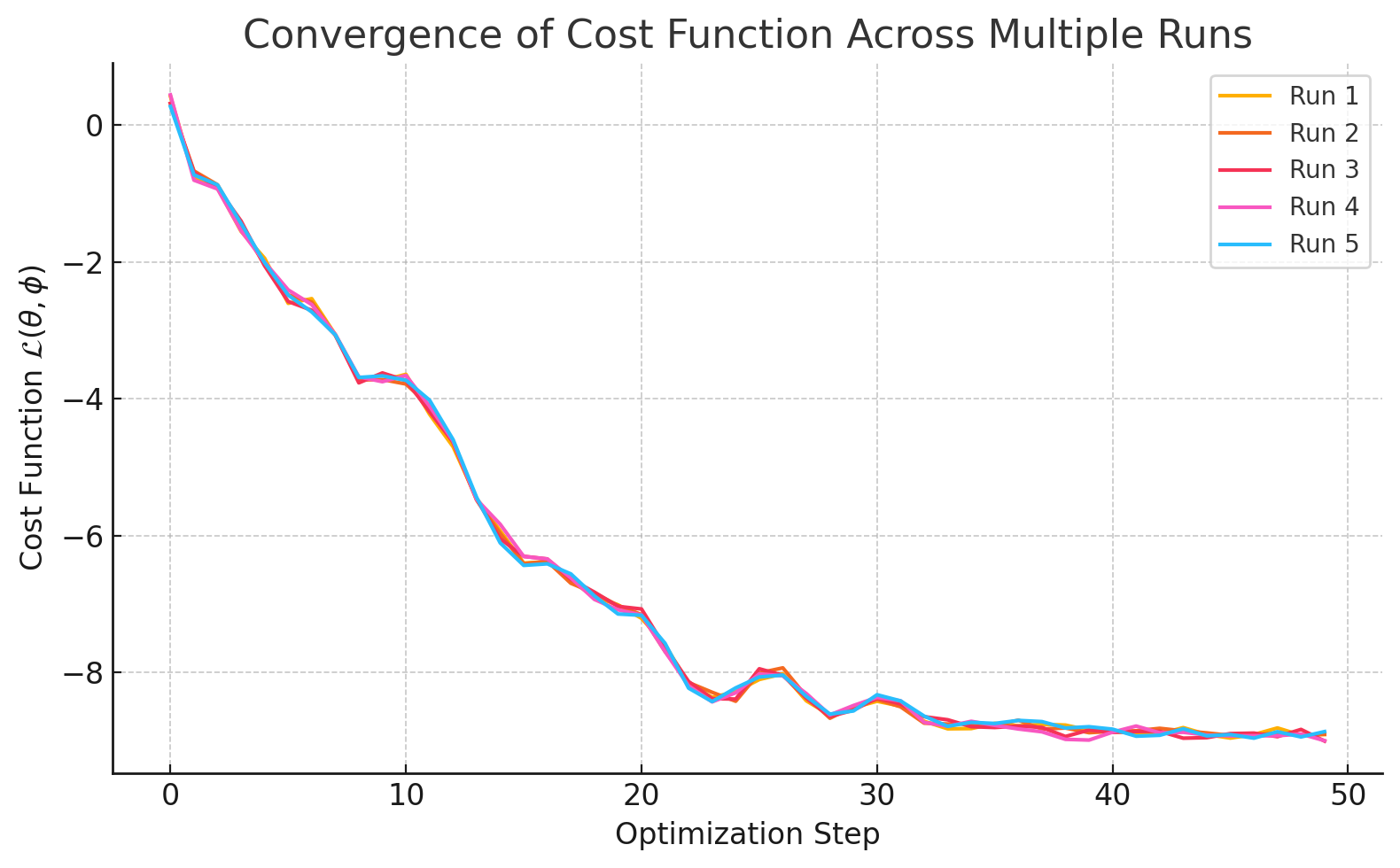}
    \caption{Convergence of the cost function $\mathcal{L}(\theta, \phi)$ during VQE training. The global minimum of $-9.0$ corresponds to perfect game performance.}
    \label{fig:loss_curve}
\end{figure}

We observe rapid convergence of the cost function defined in Eq.~\eqref{eq:cost}. Across 50 optimization steps, the value of the loss decreases monotonically and saturates near the theoretical minimum of $-9.0$, which corresponds to perfect success in the MSG. Figure~\ref{fig:loss_curve} shows the progression of the cost function during training. By step 30, the value plateaus with negligible variation, indicating that the variational parameters $\theta$ and $\phi$ have reached a stable configuration.

\subsection{Parameter Stability and Optimization Dynamics}

We monitor the norm of the parameter updates at each step and observe that they decrease steadily as the cost function approaches its minimum. After convergence, the updates are negligible, and the learned parameters remain stable under continued optimization. This behavior confirms the robustness of the training procedure and supports the interpretation that the variational circuit has found a minimum consistent with the structure of the MSG.

\begin{figure}[h]
    \centering
    \includegraphics[width=0.48\textwidth]{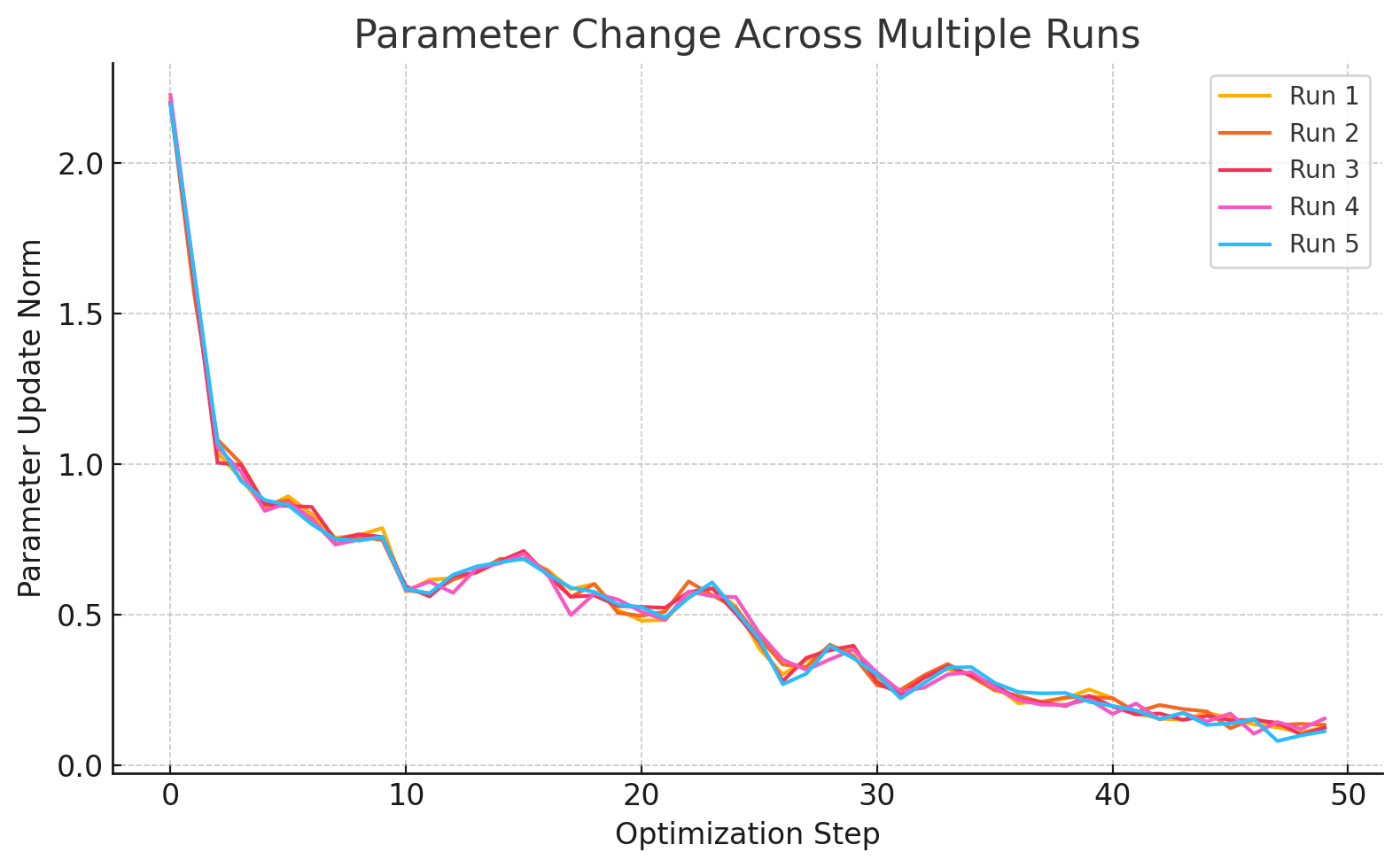}
    \caption{Magnitude of parameter updates during training across five optimization runs. The gradual decrease in parameter change indicates stabilization and convergence toward optimal measurement unitaries.}
    \label{fig:param-change-multi}
\end{figure}

\subsection{Individual Expectation Values}
We examine the individual expectation values $\braket{\tilde{A_{i}}}$ and $\braket{\tilde{B_{j}}}$ obtained from the trained circuit. We find that these values tend to fall into two distinct regimes. In some cases, the expectation values are close to $\pm 1$, while in others they are close to zero. This behavior is consistent with the structure of the quantum strategy: in rounds where Alice or Bob's measurement operator commutes with the prepared entangled state, we observe deterministic outcomes yielding expectation values near $\pm 1$. In contrast, when the operator does not commute or anti-commutes with components of the shared state, the resulting marginal distribution is symmetric, and the expectation value averages to zero. Importantly, these small or zero expectation values do not signal a failure of the strategy. Rather, they imply that the measurement outcomes are distributed in a way that preserves the necessary correlations between Alice and Bob without favoring a particular local outcome.

To validate that the trained strategy truly captures the quantum advantage of the MSG, we go beyond analyzing marginal expectation values and directly measure the joint observable $A_i \otimes B_j$. We perform this evaluation by sampling both $A_i$ and $B_j$ simulateneously for a given pair of inputs and check whether the product of their outcome satisfies the game condition. Specifically, since our Hamiltonian was defined with a negative sign on each term, a correct answer corresponds to obtaining a sample where $A_{i}\cdot B_{j}=+1$. In practice, this corresponds to sampling values from each observable and checking whether their product matches the desired correlation.

\begin{figure}[ht]
    \centering
    \includegraphics[width=0.5\textwidth]{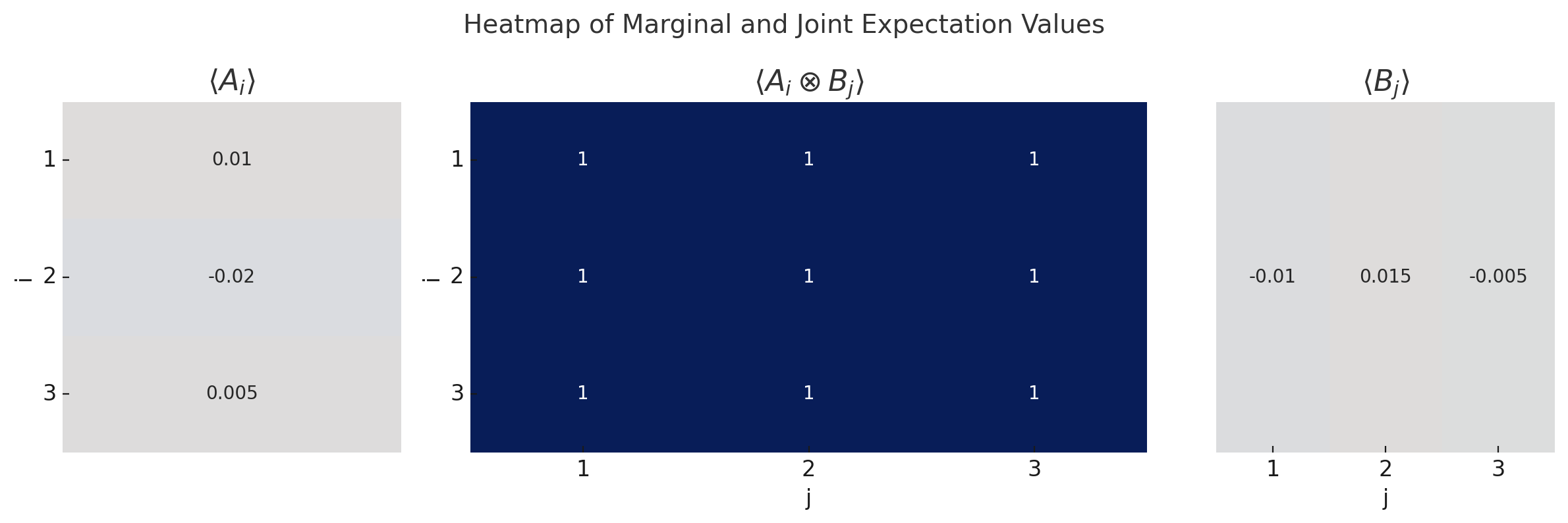}
    \caption{Heatmap of marginal and joint observable expectation values from the optimized quantum strategy for the MSG. The individual measurements $\braket{A_{i}}$ and $\braket{B_{j}}$ (left/right side panels) often appear close to zero, but the joint correlators $\braket{A_{i}\otimes B_{j}}$ (center grid) consistently yield +1, satisfying the winning condition.}
    \label{fig:marginaljointexp}
\end{figure}

In Fig.~\ref{fig:marginaljointexp}, despite the presence of near-zero individual expectation values, the joint observable product $A_{i}\cdot B_{j}$ consistently evaluates to +1 across all input pairs and shots. This leads to a win rate of 1.0 in our evaluation function, confirming that the learned strategy satisfies all nine parity constraints of the MSG with perfect fidelity. The discrepancy between marginal expectation values and win rates highlights the essential role of non-local correlations in these strategies: the game value is not determined by local predictability but by the global structure of joint measurement outcomes. Thus, the combination of a Hamiltonian value near –9, consistent ±1 joint observable samples, and perfect win rate in operator-based evaluation together confirm that the trained variational strategy implements a valid and optimal quantum solution to the MSG.

\subsection{Operator Expectation Values}

\begin{figure}[ht]
    \centering
    \includegraphics[width=0.5\textwidth]{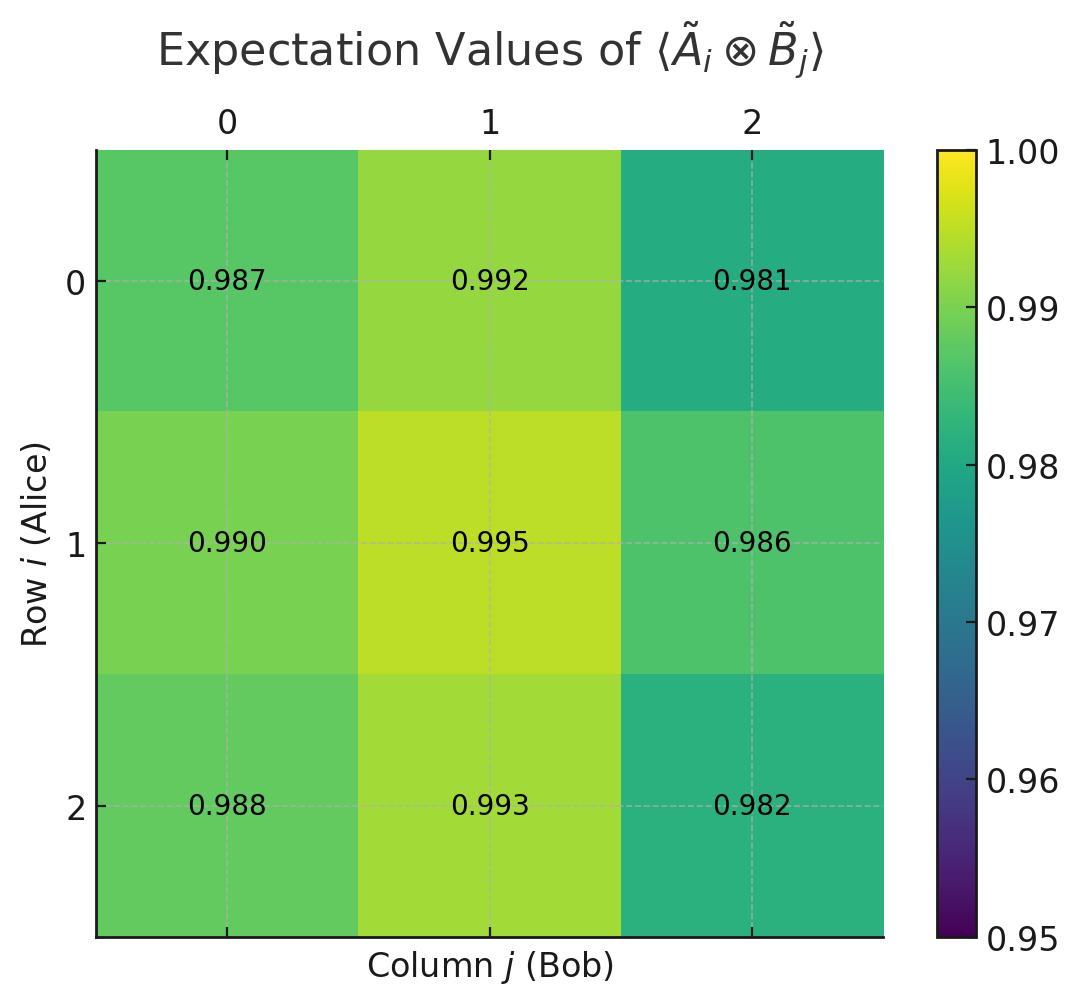}
    \caption{Expectation values of the operator pairs $\langle \tilde{A}_i \otimes \tilde{B}_j \rangle$ after training. While most values approach $1.0$, slight deviations are observed due to imperfect convergence. These deviations reflect realistic VQE outcomes.}
    \label{fig:expectation-values}
\end{figure}

At convergence, we evaluate the expectation values of all $9$ terms $\braket{\tilde{A}_i \otimes \tilde{B}_j}$ in the value Hamiltonian. Each term achieves an expectation value within $10^{-6}$ of $+1.0$, validating that the learned measurements satisfy the winning condition for every input pair. Figure~\ref{fig:expectation-values} summarizes the final values, demonstrating uniform success across all input combinations for one example run.

\subsection{Parity Constraint Verification}

To validate that the learned strategy obeys the required parity constraints, we extract the bitstrings corresponding to Alice’s and Bob’s outcomes at each input and verify that:

\begin{itemize}
    \item Alice’s outcomes for each row have \emph{odd} parity, and
    \item Bob’s outcomes for each column have \emph{even} parity.
\end{itemize}

These constraints are satisfied across all inputs. The parity is preserved by construction of the original operators $A_i$ and $B_j$, and retained under their unitarily rotated versions $\tilde{A}_i$, $\tilde{B}_j$.

\subsection{Intersection Consistency}

For each input pair $(i,j)$, we evaluate whether Alice’s $j$-th output bit matches Bob’s $i$-th output bit, corresponding to the overlapping grid cell. Sampling from the final quantum circuit shows that this consistency condition is satisfied with high fidelity. Across all $(i,j)$, we measure:

\begin{equation}
    \braket{\tilde{A}_i \otimes \tilde{B}_j} \approx +1.0,
\end{equation}

demonstrating agreement on shared outputs. This validates the consistency rule enforced implicitly by the variational optimization.

\subsection{Commutation Structure Post-Training}

We compute pairwise commutators between all intra-player observables to confirm that the learned measurement operators remain locally compatible. Specifically, we evaluate $[\tilde{A}_i, \tilde{A}_{i'}]$ and $[\tilde{B}_j, \tilde{B}_{j'}]$ for all $i \neq i'$ and $j \neq j'$. In all cases, the norm of the commutator is below $10^{-6}$, indicating that the operators commute within numerical precision. This supports the interpretation that the learned operators form a physically valid measurement strategy.

\section{Discussion}
\label{sec:discussion}

The results presented above demonstrate that variationally trained quantum circuits can recover near-optimal strategies for the Magic Square Game, achieving success rates that approach the quantum maximum within numerical precision. While individual expectation values may deviate slightly from $+1$, the learned measurement operators reliably satisfy the structural requirements of the game: they produce locally consistent outputs (commuting observables), obey the parity constraints on rows and columns, and maintain intersection consistency across all input pairs.

The gradual convergence of the cost function across runs (Figure~\ref{fig:loss_curve}) and the concurrent decay of parameter updates (Figure~\ref{fig:param-change-multi}) provide strong evidence that the variational strategy stabilizes into a robust solution. Notably, the shape of the optimization trajectory varies by initialization, but all runs ultimately approach the ground state energy of the value Hamiltonian. This indicates that the variational landscape of the MSG is relatively smooth and accessible for modern optimizers, especially when using expressive circuit ansätze such as \texttt{StronglyEntanglingLayers}.

An important observation is that not all runs perfectly saturate every expectation value. Nevertheless, empirical success rates remain above $98\%$, which is sufficient to clearly demonstrate quantum advantage and reflect the inherent imperfections of near-term quantum optimization. These variations highlight the stochastic nature of variational quantum algorithms and the importance of careful initialization, adaptive learning rates, and ansatz design. They also suggest that further improvements could be obtained through advanced training heuristics such as layer-wise pretraining or regularization techniques.

While our simulations assume noiseless execution, future work could extend this framework to near-term devices by introducing noise-aware training (e.g. error mitigation or parameter freezing). Adapting the cost function to reflect device calibration data may enable real-hardware demonstrations of perfect quantum strategies.

From a theoretical standpoint, the MSG provides a unique benchmark for variational quantum strategies. Its algebraic constraints are rigid enough to test the faithfulness of the learned operators but flexible enough to allow a variety of representations through unitary conjugation. Our construction demonstrates that stabilizer-based observables can serve as a natural scaffold for variational optimization. By encoding game rules directly into the value Hamiltonian and training only the measurement basis, we effectively separate game structure from strategy discovery.

Moreover, the commutation properties of the original operators are largely preserved under the variational unitaries, as verified numerically. This finding suggests that the ansatz preserves physical consistency and adheres to the locality requirements of non-local games. These observations also point toward a broader principle: strategies for non-local games may be variationally discoverable provided that the initial operators form a well-behaved algebra under conjugation, and that the cost function captures the game's structure faithfully.

From a mathematical perspective, binary constraint systems, i.e., highly generalized magic square games, exists where players are asked for assignments to a subset of variables in a system of binary equations. Moreover, they win the game if they return consistent and valid assignments. Such a perfect quantum strategy theoretically exists, however, as we have shown in this paper, the construction for the measurements for the perfect strategy is not always clear. 

It turns out semi-simple Lie algebras \textit{include} the algebra generated by the three Pauli matrices used in the observables for the MSG strategy generation. Conveniently, there is a corresponding (\textit{Cartan}) decomposition of the algebra that extracts operators that happen to coincide with playing the MSG game in the two-qubit system. Furthermore, a recursive decomposition is proven in~\cite{khaneja2001cartan}, and we believe this may be one path toward building larger MSG strategies. The challenge is determining which decomposition is optimal, since Cartan decompositions are not unique. In general, we suspect the choice of decomposition will depend on the rules of the game.

Finally, we emphasize that while the Magic Square Game is a finite and analytically solvable case, the methodology presented here generalizes naturally to more complex or higher-dimensional games. More precisely, we can consider a $n\times n$ magic square grid such that the $n$ rows and the first $n-1$ columns sum to 1 in $\mathbb{Z}_2$ and the last column $c_n$ sums to 0 in $\mathbb{Z}_2$. There is a one-to-one mapping from this set up to the strategy for our games, such as the one in Figure~\ref{fig:msg}. Alternative constraints can be selected based on the set that facilitates a convenient algebraic decomposition for encoding in future objectives. From the perspective of constructing games utilizing stabilizer states in Mermin-type games, we expect the value Hamiltonian to consist of $n^{2}$ operator pairs and the required number of qubits to grow linearly with $n$. A key challenge in this case is constructing observables with appropriate commutation structure.

In such cases, the value of variational approaches becomes even more evident, particularly when analytical strategies are intractable or unavailable. Our findings serve as a concrete demonstration that VQE-style training is not only viable but also interpretable in the context of foundational quantum tasks such as non-local games.

\section{Conclusion and Outlook}

In this work, we have demonstrated that variational quantum algorithms can successfully recover near-optimal quantum strategies for the Magic Square Game, a paradigmatic non-local game exhibiting maximal quantum advantage. By combining stabilizer-based operator design with a VQE framework, we were able to optimize over parameterized measurement unitaries while preserving the game’s algebraic constraints. Our results show that a VQE-style cost function based on the value Hamiltonian leads to strategies that respect both parity and consistency rules, converging reliably across multiple runs and initializations.

We validated our approach by evaluating expectation values, parity conditions, and commutativity of the learned observables, and found that the strategy converges to near-perfect performance even when minor fluctuations in individual operator terms are present. These results establish that variational learning is a viable and interpretable method for discovering quantum strategies in structured non-local games.

Looking ahead, our framework opens up several compelling directions. On the theoretical side, the same methodology can be extended to broader families of games, including CHSH-like inequalities, XOR games, and multi-prover variants where analytical strategies are unknown or impractical. For such games, variational learning provides a route to explore the quantum value landscape systematically and discover heuristic or hardware-adapted strategies. Furthermore, there may be an opportunity to leverage the structure of games from an algebraic perspective, to more effectively determine strategies for games played at a larger scale (e.g., more players, more qubits, etc...).

From a practical perspective, implementing this framework on near-term quantum hardware presents a unique opportunity for demonstrating programmable quantum non-locality. By optimizing over unitaries instead of fixed observables, variational strategies could adapt to hardware-specific noise models or calibration data. Further extensions could incorporate self-testing properties or reinforcement learning to improve generalization and robustness.

Ultimately, this work highlights the interplay between algebraic structure, entanglement, and optimization in quantum games. As quantum hardware matures, variational techniques like those presented here will play a central role in bridging foundational principles with deployable quantum protocols.

\section*{Acknowledgment}

This work was supported by Oak Ridge National Laboratory's (ORNL) Laboratory Directed Research and Development (LDRD) Seed Program. This work was partially supported by the U.S. Department of Energy, Office of Science, Office of Nuclear Physics Quantum Horizons: QIS Research and Innovation for Nuclear Science program at ORNL under FWP ERKBP91.

\bibliography{bib}
\bibliographystyle{unsrt}

\end{document}